%% file: arxiv-manuscript.tex
\documentclass{article}

\usepackage{arxiv}
\usepackage{graphicx}
\usepackage{newtxtext}
\usepackage{newtxmath}
\usepackage{natbib}
\usepackage{hyperref}
\usepackage{siunitx}
\usepackage{multirow,bigdelim}
\usepackage[utf8]{inputenc} 
\usepackage[T1]{fontenc}    
\usepackage{url}            
\usepackage{booktabs}       
\usepackage{amsfonts}       
\usepackage{nicefrac}       
\usepackage{microtype}      
\usepackage{doi}
\usepackage{bm}

\hypersetup{
    colorlinks = true,
    urlcolor   = blue,
    citecolor  = black,
}

\newcommand{\nn}{\nonumber}
\newcommand{\angles}[1]{\left\langle#1\right\rangle} 
\newcommand{\mean}[3]{\angles{\left\{#1\right\}_{#2}^{#3}}} 
\newcommand{\He}{\mathcal{H}^{\mathrm{e}}} 
\newcommand{\Hd}{\mathcal{H}^{\mathrm{d}}} 
\newcommand{\dx}{\mathrm{d}x} 

\title{Combined proper orthogonal decompositions\\of orthogonal subspaces}

\author{ \href{https://orcid.org/0000-0003-3444-493X}{\includegraphics[scale=0.06]{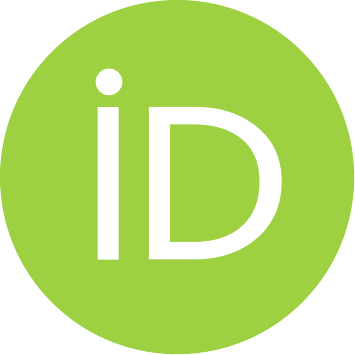}\hspace{1mm}Peder J.~Olesen}\\
	Department of Civil and Mechanical Engineering\\
	Technical University of Denmark\\
	2800 Kongens Lyngby, Denmark\\
	\texttt{pjool@dtu.dk} \\
		\And
	\href{https://orcid.org/0000-0003-1307-5290}{\includegraphics[scale=0.06]{Figures/orcid.pdf}\hspace{1mm}Azur Hod\v{z}ic} \\
	Department of Civil and Mechanical Engineering\\
	Technical University of Denmark\\
	2800 Kongens Lyngby, Denmark\\
  \texttt{azur.hod@gmail.com}
     \And
	\href{https://orcid.org/0000-0002-8657-0383}{\includegraphics[scale=0.06]{Figures/orcid.pdf}\hspace{1mm}Clara M.~Velte} \\
	Department of Civil and Mechanical Engineering\\
	Technical University of Denmark\\
	2800 Kongens Lyngby, Denmark\\
  \texttt{cmve@dtu.dk}
}

\renewcommand{\headeright}{}
\renewcommand{\undertitle}{}
\renewcommand{\shorttitle}{Combined proper orthogonal decompositions of orthogonal subspaces}

\hypersetup{
pdftitle={Combined proper orthogonal decompositions of orthogonal subspaces},
pdfsubject={phys.fluid},
pdfauthor={Peder J.~Olesen, Azur Hod\v{z}i\'c, Clara M.~Velte},
pdfkeywords={Modal decomposition, turbulence},
}

\begin{document}
\maketitle

\begin{abstract}
We present a method for combining proper orthogonal decomposition (POD) bases optimized with respect to different norms into a single complete basis. We produce a basis combining decompositions optimized with respect to turbulent kinetic energy (TKE) and dissipation rate. The method consists of projecting a data set into the subspace spanned by the lowest several TKE optimized POD modes, followed by decomposing the complementary component of the data set using dissipation optimized POD velocity modes. The method can be fine-tuned by varying the number of TKE optimized modes, and may be generalized to accommodate any combination of decompositions. We show that the combined basis reduces the degree of non-orthogonality compared to dissipation optimized velocity modes. The convergence rate of the combined modal reconstruction of the TKE production is shown to exceed that of the energy and dissipation based decompositions. This is achieved by utilizing the different spatial focuses of TKE and dissipation optimized decompositions.
\end{abstract}

\section{Introduction}
Reduced order models (ROMs) of turbulent flows approximate the high-dimensional flow dynamics by projecting them into a lower dimensional subspace spanned by a truncated modal basis. Ideally, the full basis should (1) be complete, in the sense that the full dynamics are recovered when all modes are included, and (2) ensure that the truncated model preserves critical aspects of the dynamics such that the approximation remains meaningful. 

Proper orthogonal decomposition (POD) provides a complete modal basis optimized for representing the underlying data set, minimizing the error as measured by a given norm. While this basis satisfies (1) by construction, it is not \emph{a priori} given that (2) holds \citep{holmes2012turbulence}. In canonical implementations of POD and POD-based ROMs the norm is chosen to optimally represent the mean turbulent kinetic energy (TKE) \citep{Lumley1967}. Such ROMs prioritize TKE-rich large-scale structures, at the expense of small-scale dissipative structures encoded in truncated modes, potentially leading to inaccuracies and instabilities.

Among the general approaches for enhancing ROM stability and accuracy are \emph{closure models}, in which the effects of unresolved modes are modelled by introducing artificial dissipation, and \emph{modification of the optimization problem itself} to better capture all relevant scales in the resulting POD modes \citep{bergmann2009enablers}. \citet{aubry1988dynamics} applied the former approach through an effective viscosity model; it has since been suggested that viscosity specific to modes or to pairs of interacting modes might be used instead of global viscosity corrections \citep{rempfer1994evolution,rempfer1996investigations}. More recently, \citet{wang2012proper} investigated a number of different closure models.

The second approach involves modifying the POD procedure itself. \cite{christensen1999evaluation} demonstrated a procedure with some similarities to what is shown in the present work, forming a partial basis using a decomposition of predefined states and a complementary basis using a decomposition of the data component orthogonal to the first basis. \citet{iollo2000two} formulated a Sobolev norm minimizing combined error in velocity and gradient fields. \citet{kostas2005comparison} used enstrophy-optimized vorticity modes to analyse the velocity and vorticity fields for a backward-facing step flow. Similarly, \citet{lee2020improving} used enstrophy-optimized modes for the expansion of gradient terms to supplement the classical TKE-optimized POD basis. \citet{olesen2023dissipation} presented a related dissipation optimized decomposition, and demonstrated a method for spanning the velocity field using such modes. The resulting velocity basis is complete (satisfying (1)), though it might suffer from similar issues as energy-optimized POD bases regarding (2), due to poor representation of energetic large-scale structures. Directly combining the two bases would generally compromise (1), producing either an incomplete or an overcomplete set of modes. In the present work we lay out a generalizable method for combining energy and dissipation optimized POD bases in such a way as to ensure exact completeness of the resulting basis, while also allowing for an adjustable balancing of the two optimizations.

The remainder of this paper is laid out as follows. The basic POD formalism is summarized in Section \ref{sec:pod}, and the combined POD formalism is laid out in Section \ref{sec:combined_pod}. Basic POD results are given in Section \ref{sec:implementation}. Convergence of reconstructed TKE, TKE production, and dissipation rate are investigated in Section \ref{sec:reconstruction}. A conclusion is presented in Section \ref{sec:conclusion}.

\section{Single-basis POD formalisms\label{sec:pod}}
This section summarizes the basic formalism for the TKE and dissipation rate optimized PODs (e-POD and d-POD, respectively), including the computation of d-POD velocity modes. The formalism largely follows that presented in \citet{olesen2023dissipation}.

\subsection{Energy-optimized POD\label{sec:epod}}
We consider an ensemble of flow realisations in the form of velocity fluctuation snapshots $\mathcal{U} = \left\{\bm{u}_m\right\}_{m=1}^M \subset \He$ defined on the domain $\Omega^{\mathrm{e}}$, where $\He$ is the Hilbert space defined by
\begin{align}
    \He &:= \left\{\bm{\alpha}: \Omega^{\mathrm{e}} \rightarrow\mathbb{R}^3 \left| \bm{\alpha} \in C^1\,,\, \sum_{i=1}^3 \int_{\Omega^{\mathrm{e}}} \left|\alpha^i\right|^2 \, \dx < \infty\right\}\right.\,.
\end{align}
While not necessary for performing e-POD in itself, the requirement that $\bm{\alpha}$ be differentiable ($\bm{\alpha}\in C^1$) is included here as it will be needed later when computing strain rate tensors.

This Hilbert space is equipped with the inner product $\left(\cdot,\cdot\right)_{\He}$ and the norm $\left\lVert\cdot\right\rVert_{\He}$:
\begin{align}
    \left(\bm{\alpha},\bm{\beta}\right)_{\He} &= \sum_{i=1}^3 \int_{\Omega^{\mathrm{e}}} \alpha^i \beta^i \, \dx\,,\quad \left\lVert\bm{\alpha}\right\rVert_{\He} = \sqrt{\left(\bm{\alpha},\bm{\alpha}\right)_{\He}}\,, \quad \bm{\alpha},\bm{\beta} \in \He\,.
\end{align}

We define the e-POD operator $R^{\mathrm{e}}: \He\rightarrow\He$ associated with the ensemble $\mathcal{U}$ by its action on $\bm{\alpha}\in\He$,
\begin{align}
    R^{\mathrm{e}}\bm{\alpha} &= \mean{\left(\bm{u}_m, \bm{\alpha}\right)_{\He} \bm{u}_m}{m=1}{M}\,,
    \label{eq:epod_operator}
\end{align}
where $\angles{\cdot}$ denotes the averaging operation. The e-POD operator has orthogonal eigenmodes $\left\{\bm{\varphi}_n\right\}_{n=1}^N$ (e-POD modes) and real and non-negative eigenvalues $\left\{\lambda_n^{\mathrm{e}}\right\}_{n=1}^N$ which are assumed to be indexed in descending order,
\begin{align}
    R^{\mathrm{e}} \bm{\varphi}_n &= \lambda_n^{\mathrm{e}} \bm{\varphi}_n\,, \quad \left(\bm{\varphi}_n, \bm{\varphi}_{n'}\right)_{\He} = \delta_{nn'}\,, \quad \lambda_1^{\mathrm{e}} \geq \lambda_2^{\mathrm{e}} \geq \ldots \geq \lambda_N^{\mathrm{e}} \geq 0\,,
    \label{eq:epod_evp}
\end{align}
where $\delta_{nn'}$ is the Kronecker delta.

The e-POD modes form a complete orthogonal basis for $\mathcal{U}$, allowing flow realisations to be expanded using uncorrelated coefficients $\left\{a_{mn}\right\}_{n=1}^N$,
\begin{align}
    \bm{u}_m &= \sum_{n=1}^N a_{mn}\bm{\varphi}_n\,,\quad a_{mn} = \left(\bm{\varphi}_n, \bm{u}_m\right)_{\He}\,, \quad \mean{a_{mn}a_{mn'}}{m=1}{M} = \lambda_{n}^{\mathrm{e}}\delta_{nn'}\,.
    \label{eq:epod_expansion}
\end{align}

This expansion is optimal with respect to $\left\lVert\cdot\right\rVert_{\He}$, in the sense that truncating the expansion in \eqref{eq:epod_expansion} to $\hat{N} \leq N$ terms minimizes the ensemble mean error as measured by $\left\lVert\cdot\right\rVert_{\He}$, compared to any other $\hat{N}$-term expansion. Since the square of this norm is proportional to TKE, the expansion provides the most efficient modal reconstruction of TKE possible. The lowest e-POD modes represent the flow structures carrying the most TKE in the mean, corresponding in general to structures residing on the largest scales where most of the TKE resides.

\subsection{Dissipation optimized POD\label{sec:dpod}}
\citet{olesen2023dissipation} developed an analogous formalism leading to a d-POD basis that spans the corresponding ensemble of strain rate tensors (SRTs) $\mathcal{S} = \left\{\bm{s}_m\right\}_{m=1}^M\subset \Hd$, defined on the domain $\Omega^{\mathrm{d}}$. This SRT basis can then be mapped to a dissipation optimized velocity basis using a spectral inverse SRT operator. The Hilbert space containing $\mathcal{S}$ is
\begin{align}
    \Hd &:= \left\{\bm{\alpha}: \Omega^{\mathrm{d}}\rightarrow \mathbb{R}^{3\times 3} \left| \sum_{i,j=1}^3 \int_{\Omega^{\mathrm{d}}} \left|\alpha^{ij}\right|^2 \, \dx < \infty\right\}\right.\,,
    \label{eq:dpod_hilbert}
\end{align}
which is equipped with the inner product $(\cdot,\cdot)_{\Hd}$ and norm $\left\lVert\cdot\right\rVert_{\Hd}$ given by
\begin{align}
    \left(\bm{\alpha},\bm{\beta}\right)_{\Hd} &= \sum_{i,j=1}^3 \int_{\Omega^{\mathrm{d}}} \alpha^{ij} \beta^{ij} \, \dx\,, \quad \left\lVert\bm{\alpha}\right\rVert_{\Hd} = \sqrt{\left(\bm{\alpha},\bm{\alpha}\right)_{\Hd}}\,; \quad \bm{\alpha}, \bm{\beta} \in \Hd\,.
\end{align}

The SRT snapshots forming the ensemble $\mathcal{S}$ are derived from velocity fluctuation snapshots in $\mathcal{U}$ using the SRT operator $D:\, \He \rightarrow \Hd$,
\begin{align}
    \bm{s}_m &= D \bm{u}_m\,,\quad \left(D \bm{\alpha}\right)^{ij} = \frac{1}{2}\left(\nabla^i \alpha^j + \nabla^j \alpha^i\right)\,,\quad \bm{\alpha} \in \He\,.
    \label{eq:srt_operator}
\end{align}

The d-POD operator $R^{\mathrm{d}}$ and the corresponding eigenvalue problem are then formed in analogy with \eqref{eq:epod_operator} and \eqref{eq:epod_evp}, replacing $\mathcal{H}^{\mathrm{e}}$ with $\mathcal{H}^{\mathrm{d}}$ and $\mathcal{U}$ with $\mathcal{S}$:
\begin{subequations}
    \begin{align}
        R^{\mathrm{d}}\bm{\alpha} &= \mean{\left(\bm{s}_m, \bm{\alpha}\right)_{\Hd} \bm{s}_m}{m=1}{M}\,, \quad \bm{\alpha} \in \mathcal{H}^{\mathrm{d}}\,;
        \label{eq:dpod_operator}\\
        R^{\mathrm{d}} \bm{\psi}_n &= \lambda_n^{\mathrm{d}}\bm{\psi}_n\,, \quad \left(\bm{\psi}_n, \bm{\psi}_{n'}\right)_{\Hd} = \delta_{nn'}\,,\quad \lambda_1^{\mathrm{d}} \geq \lambda_2^{\mathrm{d}} \geq \ldots \geq \lambda_N^{\mathrm{d}} \geq 0\,.
        \label{eq:dpod_evp}
    \end{align}
\end{subequations}

The eigenmodes $\{\bm{\psi}_n\}_{n=1}^N \subset \mathcal{H}^{\mathrm{d}}$ form the d-POD basis. SRTs $\bm{s}_m \in \mathcal{S}$ may now be expanded in this basis, again with uncorrelated coefficients $\left\{b_{mn}\right\}_{n=1}^N$,
\begin{align}
    \bm{s}_m &= \sum_{n=1}^N b_{mn} \bm{\psi}_n\,,\quad b_{mn} = \left(\bm{\psi}_n, \bm{s}_m\right)_{\Hd}\,,\quad \mean{b_{mn} b_{mn'}}{m=1}{M} = \lambda_{n}^{\mathrm{d}}\delta_{nn'}\,.
    \label{eq:dpod_expansion}
\end{align}

This expansion is optimal with respect to $\left\lVert\cdot\right\rVert_{\Hd}$ in a sense analogous to that discussed above for the e-POD. Since the square of this norm is proportional to the mean dissipation rate, the resulting basis gives a reconstruction of the SRT which is optimal with respect to the dissipation rate. While dissipation is associated with small scales in the flow, structures associated with d-POD modes have been found to span a range of scales throughout the d-POD spectrum, as discussed by \citet{olesen2023dissipation}.

\citet{olesen2023dissipation} formulated a spectral inverse SRT operator $D^{-1}: \Hd \rightarrow \He$, using the one-to-one correspondence given by \eqref{eq:srt_operator} between fluctuation velocity snapshots $\bm{u}_m \in \mathcal{U}$ and SRT snapshots $\bm{s}_m\in\mathcal{S}$. We have for any $\bm{\psi}_n$ with $\lambda_n^{\mathrm{d}}\neq 0$
\begin{align}
    D^{-1} \bm{\psi}_n &= \frac{1}{\lambda_n^{\mathrm{d}}}\mean{\left(\bm{s}_m, \bm{\psi}_n\right)_{\Hd} \bm{u}_m}{m=1}{M}\,.
    \label{eq:srt_inv_operator}
\end{align}
This operation produces a velocity field corresponding to each d-POD mode with $\lambda_n^{\mathrm{d}} \neq 0$, and the resulting set of d-POD velocity fields $\left\{D^{-1} \bm{\psi}_n\right\}_{\lambda_n^{\mathrm{d}}\neq 0} \subset \mathcal{H}^{\mathrm{e}}$ forms a complete basis for $\mathcal{U}$. The optimality with respect to dissipation is inherited by this basis, meaning that any velocity-derived term can be expanded in a dissipation-optimized manner.

\section{Combined POD bases\label{sec:combined_pod}}
The formalism laid out in Section \ref{sec:pod} results in two distinct velocity bases for $\mathcal{U}$. The e-POD basis is optimal with respect to TKE, and provides an efficient representation of large-scale turbulent structures in the flow. It possesses all of the attractive properties associated with POD, including modal orthogonality and uncorrelated coefficients \eqref{eq:epod_expansion}. For a channel flow, the e-POD basis is particularly efficient for reconstructing flow features located in the TKE-rich bulk region, while it is less efficient for features in the near-wall region \citep{olesen2023dissipation}. The d-POD velocity basis, on the other hand, gives a dissipation rate optimized reconstruction. It reconstructs near-wall features in a turbulent channel flow more efficiently than does e-POD, including the TKE density in this region. Like the case of e-POD, the d-POD velocity basis is characterized by uncorrelated expansion coefficients, but it is in general not orthogonal with respect to $\left(\cdot,\cdot\right)_{\He}$.

In this section we present a method for combining the two velocity bases so as to balance the representation of energetic and dissipative flow features. The idea is to project the flow onto the lowest $N'$ e-POD modes, perform a complementary d-POD on the unresolved part of the flow, and map the d-POD modes to velocity modes using \eqref{eq:srt_inv_operator}. A complete basis for the velocity data set is formed by combining the $N'$ e-POD modes with the complementary d-POD velocity basis.

First, e-POD is performed on $\mathcal{U}$ as shown in Section \ref{sec:epod}. We then project $\mathcal{U}$ into the subspace orthogonal to the first $N' \leq N$ e-POD modes, resulting in the projected data set $\mathcal{U}^{\perp N'} = \left\{\bm{u}_m^{\perp N'}\right\}_{m=1}^M$ given by
\begin{align}
    \bm{u}_m^{\perp N'} &= \bm{u}_m - \sum_{n=1}^{N'} a_{mn} \bm{\varphi}_n\,.
    \label{eq:complementary_projection}
\end{align}

This operation represents a projection into a space of lower dimension, and thus a reduction of the rank of the data set by $N'$. SRTs are computed from $\mathcal{U}^{\perp N'}$ using \eqref{eq:srt_operator} to produce $\mathcal{S}^{\perp N'} = \left\{\bm{s}_m^{\perp N'}\right\}_{m=1}^M$. We apply d-POD as described in Section \ref{sec:dpod} to this data set, resulting in complementary d-POD modes $\{\bm{\psi}_n^{\perp N'}\}_{n=1}^{N}$ and eigenvalues $\{\lambda_n^{\mathrm{d},\perp N'}\}_{n=1}^{N}$. Due to the reduction of rank from the projection \eqref{eq:complementary_projection}, $\lambda_n^{\mathrm{d},\perp N'} = 0$ for $n > N-N'$. The complementary modes are converted to velocity fields using \eqref{eq:srt_inv_operator}, yielding a basis $\{D^{-1}\bm{\psi}^{\perp N'}_n\}_{\lambda_{n}^{\mathrm{d},\perp N'}\neq 0}$ for $\mathcal{U}^{\perp N'}$,
\begin{align}
    \bm{u}_m^{\perp N'} &= \sum_{n \left|\lambda^{\mathrm{d},\perp N'} \neq 0\right.} b_{mn}^{\perp N'} D^{-1} \bm{\psi}_n^{\perp N'}\,,\quad b_{mn}^{\perp N'} = \left(\bm{\psi}_n^{\perp N'}, \bm{s}_m^{\perp N'}\right)_{\Hd}\,.
\end{align}

Combining the complementary d-POD velocity basis with the e-POD modes subtracted in \eqref{eq:complementary_projection} produces a complete basis $\mathcal{B} = \left\{\bm{\varphi}_n\right\}_{n=1}^{N'} \cup \left\{D^{-1}\bm{\psi}_n^{\perp N'}\right\}_{\lambda_n^{\mathrm{d},\perp N'}\neq 0}$ for $\mathcal{U}$, the coefficients of which can be shown to be uncorrelated:
\begin{align}
    \bm{u}_m &= \sum_{n=1}^{N'} a_{mn} \bm{\varphi}_n + \sum_{n\left|\lambda_n^{\mathrm{d},\perp N'}\neq 0\right.} b_{mn}^{\perp N'} D^{-1}\bm{\psi}_{n}^{\perp N'}\,; \quad \mean{a_{mn}b_{mn'}^{\perp N'}}{m=1}{M} = 0\,, \quad n \leq N'\,.
    \label{eq:cpod_expansion}
\end{align}
The uncorrelatedness of coefficients implies that cross terms vanish in the expansion of second-order mean quantities.

The formalism presented here provides a gradual transition between full e-POD ($N'=N$) and full d-POD bases ($N'=0$). While the method is in principle straightforward, it hinges entirely on the use of \eqref{eq:srt_inv_operator} to map SRT modes to velocity modes. By adapting \eqref{eq:srt_inv_operator} the method can be generalized, such that any combination of decompositions may in principle be smoothly linked using the ideas presented in this section.

\section{Basic POD results\label{sec:implementation}}
In Sections \ref{sec:implementation} and \ref{sec:reconstruction} we apply the combined POD to the turbulent channel data set described in \citet{olesen2023dissipation}. The data set consists of $N=1078$ velocity fluctuation snapshots of a channel cross section obtained from a direct numerical simulation. The rank of the data set is 1077, since the subtraction of the mean field reduces the rank by one; thus, all but one eigenvalue is non-zero. For further details on the simulation and the data set we refer to the above referenced paper.

\subsection{POD spectra}
\begin{figure}
    \centerline{\input{Figures/filtered_spectra}}
    \caption{($a$): e-POD spectrum normalized by $\sum_n \lambda_n^{\mathrm{e}}$. ($b$): Complementary d-POD Spectra using e-POD subspace dimensions $N' \in \{0, 10, 50, 100, 200\}$, with $N'=0$ corresponding to the full d-POD spectrum. Each spectrum is shifted horizontally by $N'$ to align the far ends, and normalized by $\sum_n \lambda_n^{\mathrm{d},\perp 0}$.\label{fig:fil_spectra}}
\end{figure}
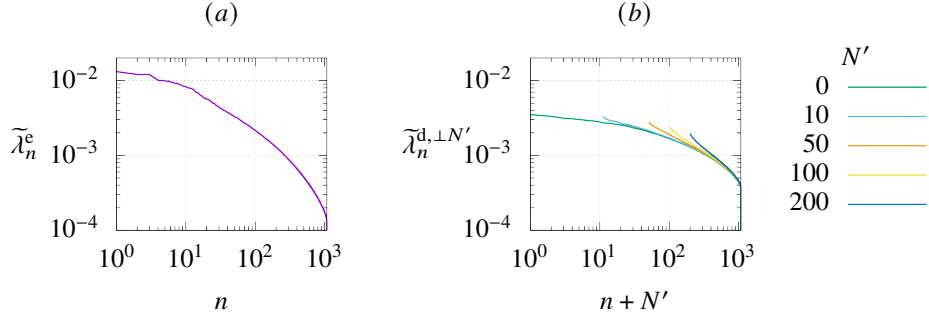

The e-POD spectrum is shown in figure \ref{fig:fil_spectra}$a$, and the d-POD spectra resulting from applying the procedure to the data set with $N' \in \{0, 10, 50, 100, 200\}$ are shown in figure \ref{fig:fil_spectra}$b$. The latter spectra are shown with indices shifted by $N'$ to align the spectra at high mode numbers. Compared to the base d-POD spectrum ($N'=0$) each of the complementary d-POD spectra ($N'>0$) is lifted for small $n$ before collapsing with increasing $n$, indicating low-dimensional dissipative structures unresolved by the e-POD sub-basis being effectively resolved by relatively few complementary d-POD modes.

\subsection{Non-orthogonality of complementary d-POD velocity modes\label{sec:overlaps}}
When building a ROM from a non-orthogonal basis, the projected system of equations is complicated by the inclusion of cross-terms. It is therefore of interest to minimize the degree of non-orthogonality when possible. While e-POD modes are orthogonal with respect to the inner product $\left(\cdot,\cdot\right)_{\He}$, and d-POD modes are orthogonal with respect to $\left(\cdot,\cdot\right)_{\Hd}$, d-POD velocity modes are generally not orthogonal to each other with respect to $\left(\cdot,\cdot\right)_{\He}$. Here we quantify the degree of non-orthogonality. We define the normalized velocity mode overlap $\epsilon^{\perp N'}_{nn'}$ as
\begin{align}
    \epsilon_{nn'}^{\perp N'} &= \frac{\left|\left(D^{-1}\bm{\psi}_n^{\perp N'}, D^{-1}\bm{\psi}_{n'}^{\perp N'}\right)_{\He}\right|}{\left\lVert D^{-1}\bm{\psi}_{n}^{\perp N'}\right\rVert_{\He} \left\lVert D^{-1}\bm{\psi}_{n'}^{\perp N'}\right\rVert_{\He}}\,.
    \label{eq:complementary_overlaps}
\end{align}

Table \ref{tab:crossterm_stats} summarizes the maximum and mean off-diagonal overlaps for pairs formed from the lowest 500 d-POD velocity modes for each $N'\in\{0,10,50,100,200\}$. The maximum magnitude of overlaps decreases when increasing $N'$ up to 100, and increases slightly at $N'=200$. Meanwhile, the mean magnitude decreases uniformly across the values of $N'$ considered. The overall trend is thus that larger e-POD subspace dimensionality leads to smaller overlaps among complementary d-POD velocity modes.

\begin{table}
    \begin{center}
        \begin{tabular}{c|c c c c c r}
             $N'$ & 0 & 10 & 50 & 100 & 200 & \\ \cline{1-6}
             $\max_{n\neq n'} \epsilon^{\perp N'}_{nn'}$ & 23.8 & 13 & 9.1 & 7.8 & 8.3 & \rdelim\}{2}{*}[${}\times 10^{-2}$]\\
             $\mean{\epsilon^{\perp N'}_{nn'}}{n\neq n'}{}$ & 3.3 & 2.3 & 1.7 & 1.4 & 1.2 & 
        \end{tabular}
        \caption{Maximum and mean off-diagonal mode overlap magnitudes $\epsilon_{nn'}^{\perp N'}$ computed from pairs formed by the lowest 500 full ($N'=0$) or complementary ($N'>0$) d-POD modes.\label{tab:crossterm_stats}}
    \end{center}
\end{table}

For an expansion including a given number of modes of a combined decomposition, the degree of non-orthogonality
compared to the expansion with the same number of full d-POD modes is ameliorated on two accounts. First, e-POD modes are orthogonal to each other as well as to complementary d-POD modes, providing orthogonality to the e-POD diagonal block as well as the off-diagonal blocks of the overlap matrix. Second, as shown here the magnitude of overlaps in the remaining complementary d-POD diagonal block are smaller. Both effects are generally enhanced when increasing e-POD subspace dimensionality, but must be balanced with the advantages of mixing the bases which will be shown in Section \ref{sec:reconstruction}.

\section{Reconstruction of TKE, TKE production, and dissipation rate\label{sec:reconstruction}}
In this section we investigate the convergence of mean TKE $\angles{T}$, TKE production $\angles{\mathcal{P}}$, and dissipation rate $\angles{\varepsilon}$ when expanded in the combined basis. Applying the expansion in \eqref{eq:cpod_expansion} leads to the following expressions, where superscripts $(1)$ and $(2)$ denote the streamwise and transverse directions in the channel, respectively, and $\nabla^{(2)} U^{(1)}$ is the transverse gradient of the streamwise mean velocity:
\begin{subequations}
    \begin{align}
        \angles{T} &= \frac{1}{2}\mean{\left|\bm{u}_m\right|^2}{m=1}{M} =  \frac{1}{2}\left(\sum_{n=1}^{N'} \lambda_n^{\mathrm{e}} \left|\bm{\varphi}_n\right|^2 + \sum_{n'\left| \lambda_{n'}^{\mathrm{d},\perp N'} \neq 0 \right.} \lambda_{n'}^{\mathrm{d},\perp N'} \left|D^{-1}\bm{\psi}_{n'}^{\perp N'}\right|^2\right)\,, 
        \label{eq:tke_reconstruct}
        \\
        \angles{\mathcal{P}} &= -\mean{u_m^{(1)} u_m^{(2)}}{m=1}{M} \nabla^{(2)} U^{(1)}\nn\\
        & = -\left(\sum_{n=1}^{N'} \lambda_n^{\mathrm{e}} \varphi_n^{(1)} \varphi_n^{(2)} + \sum_{n'\left| \lambda_{n'}^{\mathrm{d},\perp N'} \neq 0 \right.} \lambda_{n'}^{\mathrm{d},\perp N'} \left(D^{-1}\bm{\psi}_{n'}^{\perp N'}\right)^{(1)}\left(D^{-1}\bm{\psi}_{n'}^{\perp N'}\right)^{(2)} \right) \nabla^{(2)} U^{(1)}\,,
        \label{eq:prod_reconstruct}
        \\
        \angles{\varepsilon} &= 2\nu \mean{\left|D \bm{u}_m\right|^2}{m=1}{M} = 2\nu \left(\sum_{n=1}^{N'} \lambda_n^{\mathrm{e}} \left|D\bm{\varphi}_n\right|^2  + \sum_{n'=1}^{N-N'} \lambda_{n'}^{\mathrm{d},\perp N'} \left|\bm{\psi}_{n'}^{\perp N'}\right|^2\right)\,,
        \label{eq:diss_reconstruct}
    \end{align}
    \label{eq:reconstructions}
\end{subequations}
Due to the geometry of the channel flow only one component enters the mean TKE production.

Reconstructions including $\hat{N} \leq N'$ modes are built from the first $\hat{N}$ e-POD modes, and are identical to pure e-POD reconstructions, while those including $N' < \hat{N} \leq N$ include $N'$ e-POD modes and the first $\hat{N}-N'$ complementary d-POD velocity modes.

\subsection{Convergence of integrated quantities\label{sec:convergence_integrated}}
\begin{figure}
    \centerline{\input{Figures/convergence_plots}}
    \caption{Convergence of integrated mean TKE ($a$), TKE production ($b$), and dissipation rate ($c$) using $N'\in\{0, 10, 50, 100, 200,N\}$; each is normalized by their fully converged value. Convergence lead over full d-POD for integrated mean TKE ($d$) and TKE production ($e$), and over full e-POD for integrated mean dissipation rate ($f$).
    \label{fig:convergence}}
\end{figure}
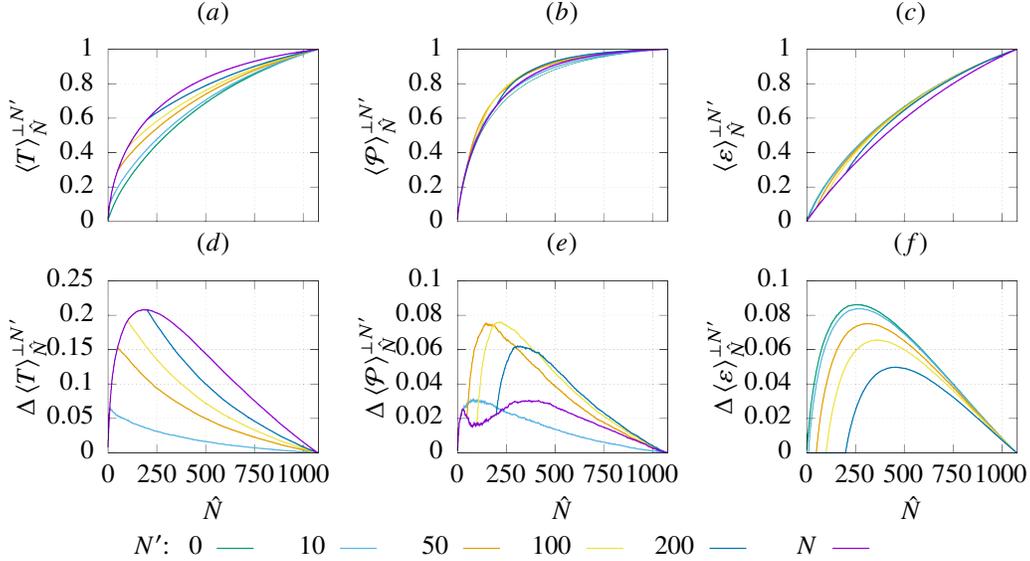

We reconstruct TKE using \eqref{eq:tke_reconstruct} with $N'\in \{0, 10, 50, 100, 200, N\}$. Figure \ref{fig:convergence}$a$ shows the convergence of the integrated mean TKE profile $\angles{T}_{\hat{N}}^{\perp N'}$, normalized to the full mean TKE. We achieve the most efficient TKE reconstruction with the full e-POD ($N'=N$), which by construction is the optimal basis for this purpose, while the full d-POD reconstruction ($N'=0$) is least effective among those considered. We define the convergence lead over the full d-POD reconstruction for each of the remaining reconstructions as
\begin{align}
    \Delta \angles{T}_{\hat{N}}^{\perp N'} &= \angles{T}_{\hat{N}}^{\perp N'} - \angles{T}_{\hat{N}}^{\perp 0}\,, \quad N'\in\left\{10,50,100,200,N\right\}\,,
    \label{eq:tke_convergence_leads}
\end{align}
which is shown in figure \ref{fig:convergence}$d$. The combined bases ($0<N'<N$) each follow the e-POD convergence for $\hat{N}\leq N'$, at which point they start to approach the d-POD curve, with the reconstructed TKE fraction remaining between that of e-POD and d-POD until $\hat{N}=N$.

For the mean TKE production $\angles{\mathcal{P}}_{\hat{N}}^{\perp N'}$, shown in figure \ref{fig:convergence}$b$, there is little difference between the convergence using full e-POD and full d-POD, although the latter is consistently the least efficient. However, in this case, the combined basis reconstruction convergence curves are not confined to the region between the full e-POD and d-POD curves. Figure \ref{fig:convergence}$e$ shows $\Delta \angles{\mathcal{P}}_{\hat{N}}^{\perp N'}$, the production convergence lead over full d-POD defined similarly to $\Delta \angles{T}_{\hat{N}}^{\perp N'}$ in \eqref{eq:tke_convergence_leads}. The combined bases each follow e-POD convergence up to $\hat{N}=N'$ as before, after which they overtake the e-POD convergence. The convergence of $\angles{\mathcal{P}}_{\hat{N}}^{\perp 10}$ falls behind that of $\angles{\mathcal{P}}_{\hat{N}}^{\perp N}$ around $\hat{N} \approx 230$, while the remaining ones maintain their leads up to the full reconstruction at $\hat{N}=N$. The maximum leads over full d-POD convergence are achieved with $N' = 50$ (at $\hat{N} \approx 140$) and $N' = 100$ (at $\hat{N} \approx 220$). While the leads are minor in absolute terms (at least in the present flow) it is nevertheless interesting to note that combining e-POD and d-POD permits a more efficient reconstruction of TKE production than either of these decompositions alone.

The mean dissipation rate reconstruction, $\angles{\varepsilon}_{\hat{N}}^{\perp N'}$, is shown in figure \ref{fig:convergence}$c$. The optimal convergence is achieved with full d-POD (which is optimized for this task), while full e-POD is the least optimal among those considered. Again, the differences are modest; figure \ref{fig:convergence}$f$ shows the dissipation convergence lead over e-POD, $\Delta\angles{\varepsilon}_{\hat{N}}^{\perp N'}$, again defined similarly to $\Delta \angles{T}_{\hat{N}}^{\perp N'}$ in \eqref{eq:tke_convergence_leads}, for $N'\in \{0,10,50,100,200\}$. Again, the convergence of each combined base follows that of e-POD up to $\hat{N}=N'$, and it remains below that of d-POD until the full reconstruction is achieved. The combined basis reconstruction convergence leads over full e-POD decreases uniformly with $N'$, while the number of modes at which the max lead occurs increases with $N'$. The behaviours observed in figures \ref{fig:convergence}$d$ and \ref{fig:convergence}$f$ demonstrate that the combined bases bridge the gap between full e-POD and d-POD, while figure \ref{fig:convergence}$e$ shows that they provide improved convergence for TKE production.

\subsection{Convergence of reconstructed profiles\label{sec:profile_reconstruction}}
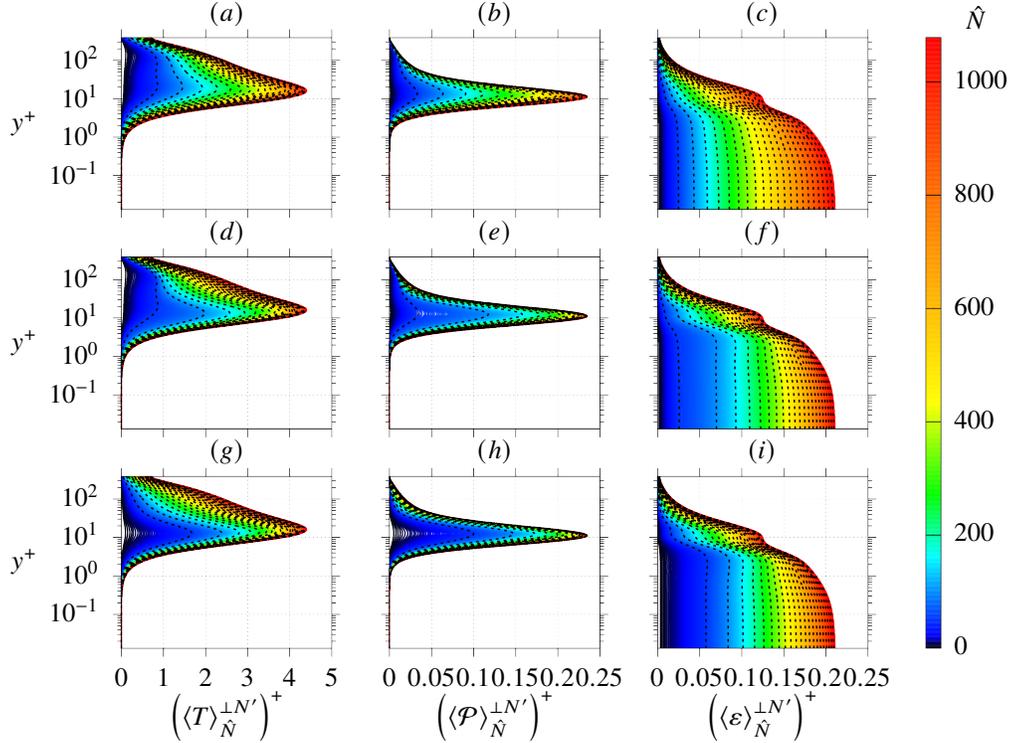
\begin{figure}
    \centerline{\input{Figures/all_cascade_DNSinit64amfNEW9-12dt2_c3}}
    \caption{Modal reconstruction of TKE (left column), TKE production (middle column), and dissipation rate (right column), using full e-POD ($a$--$c$), the combined basis with $N'=50$ ($d$--$f$), and full d-POD ($g$--$i$). All quantities are normalized to dimensionless form using kinematic viscosity $\nu$ and friction velocity $u_{\tau}$. The colour of each profile indicates the number of modes $\hat{N}$ entering the reconstruction, and dashed lines marks the profile for each additional 50 modes included.\label{fig:incremental_recon}}
\end{figure}

We reconstruct profiles of mean TKE, TKE production and dissipation rate using \eqref{eq:reconstructions} for full e-POD, full d-POD, and for the combined basis with $N'=50$. These reconstructions are shown in figure \ref{fig:incremental_recon}. As shown by \citet{olesen2023dissipation}, the full e-POD emphasizes bulk structures in the turbulent channel flow, while the full d-POD instead emphasizes near-wall structures. This is reflected in the different spatial distributions seen for low values of $\hat{N}$ in figures \ref{fig:incremental_recon}$a$ and \ref{fig:incremental_recon}$g$ for TKE, and in Figures \ref{fig:incremental_recon}$c$ and \ref{fig:incremental_recon}$i$ for dissipation rate.

The mean TKE profile reconstructed using $N'=50$, shown in figure \ref{fig:incremental_recon}$d$, combines the different spatial emphases of e-POD and d-POD. The $\hat{N}=100$ profile reconstruction, including $N'=50$ e-POD modes and $\hat{N} - N' = 50$ complementary d-POD modes, includes a significant portion of both bulk TKE (compared to the corresponding d-POD profile) and near-wall TKE (compared to the e-POD profile).

The different spatial emphases are also seen for the production profiles in figures \ref{fig:incremental_recon}$b$ and \ref{fig:incremental_recon}$h$, where the e-POD production profile for $\hat{N} = 100$ exhibits a thick tail extending into the bulk, whereas the corresponding d-POD profile captures a comparatively large part of the peak at $y^+ \approx 10$. These features are again combined for $N'=50$, for which the profile at $\hat{N}=100$ includes both features. Since much of production is localized on the transition between the dissipative near-wall region and the TKE-rich bulk, it benefits from combining the basic decompositions. This also causes the enhanced convergence of integrated TKE production using combined bases as found in Section \ref{sec:convergence_integrated}.

The mean dissipation rate is strongly localized in the near-wall region. Compared to the full d-POD reconstruction (figure \ref{fig:incremental_recon}$i$) this limits the advantage gained by enhancing the representation of the bulk region through inclusion of e-POD modes in the combined profile reconstruction (figure \ref{fig:incremental_recon}$f$).

\section{Conclusion\label{sec:conclusion}}
We have presented a method for combining different POD bases into a single complete basis, allowing for combined optimization of the overall basis which is controllable through a single parameter. We show that the combined basis reduces the magnitude of modal overlaps between complementary dissipation rate optimized velocity modes, potentially reducing the importance of cross terms in modally projected equations.

We reconstruct mean profiles of TKE, TKE production, and dissipation rate using combined TKE and dissipation rate optimized PODs for different values of the weighting parameter. The full decompositions optimized for TKE and dissipation rate show the fastest convergence for their respective optimized quantities by construction, but using combined bases for reconstructing mean TKE production yields faster convergence globally than using either of the full POD bases. This is due to balancing of different spatial emphases of the respective full POD bases.

Combining e-POD and d-POD in this way in principle gives a way to fine-tune the balance between the reconstruction of different quantities. The observed effects are likely to depend strongly on the investigated flow. In higher Reynolds number flows or less homogeneous flows we expect the quantities to be characterized by more well-separated scales and distinct structures, leading to a larger gap between the performance of e-POD and d-POD as well as a greater potential advantage from combining the two.

These features suggest that the method may lead to improved ROM performance compared to existing POD bases. It is of interest to test its capability in a full ROM, as well as its behaviour for different flows and with different combinations of POD bases.

\paragraph{Acknowledgements}The authors gratefully acknowledge the computational and data resources provided on the Sophia HPC Cluster at the Technical University of Denmark, DOI: 10.57940/FAFC-6M81.

\paragraph{Funding}PJO acknowledges financial support from the Poul Due Jensen Foundation: Financial support from the Poul Due Jensen Foundation (Grundfos Foundation) for this research is gratefully acknowledged. AH and CMV acknowledge financial support from the European Research Council: This project has received funding from the European Research Council (ERC) under the European Union's Horizon 2020 research and innovation programme (grant agreement No 803419).

\paragraph{Declaration of interests}The authors report no conflict of interest.

\paragraph{Author ORCIDs}
P.J.\ Olesen: \href{https://orcid.org/0000-0003-3444-493X}{https://orcid.org/0000-0003-3444-493X}; 
A. Hod\v{z}i\'{c}: \href{https://orcid.org/0000-0003-1307-5290}{https://orcid.org/0000-0003-1307-5290}; 
C.M.\ Velte: \href{https://orcid.org/0000-0002-8657-0383}{https://orcid.org/0000-0002-8657-0383}.

\bibliographystyle{jfm}
\bibliography{arxiv-manuscript}

\end{document}

%% file: Figures/filtered_spectra.tex
\begingroup
  \makeatletter
  \providecommand\color[2][]{%
    \GenericError{(gnuplot) \space\space\space\@spaces}{%
      Package color not loaded in conjunction with
      terminal option `colourtext'%
    }{See the gnuplot documentation for explanation.%
    }{Either use 'blacktext' in gnuplot or load the package
      color.sty in LaTeX.}%
    \renewcommand\color[2][]{}%
  }%
  \providecommand\includegraphics[2][]{%
    \GenericError{(gnuplot) \space\space\space\@spaces}{%
      Package graphicx or graphics not loaded%
    }{See the gnuplot documentation for explanation.%
    }{The gnuplot epslatex terminal needs graphicx.sty or graphics.sty.}%
    \renewcommand\includegraphics[2][]{}%
  }%
  \providecommand\rotatebox[2]{#2}%
  \@ifundefined{ifGPcolor}{%
    \newif\ifGPcolor
    \GPcolortrue
  }{}%
  \@ifundefined{ifGPblacktext}{%
    \newif\ifGPblacktext
    \GPblacktexttrue
  }{}%
  \let\gplgaddtomacro\g@addto@macro
  \gdef\gplbacktext{}%
  \gdef\gplfronttext{}%
  \makeatother
  \ifGPblacktext
    \def\colorrgb#1{}%
    \def\colorgray#1{}%
  \else
    \ifGPcolor
      \def\colorrgb#1{\color[rgb]{#1}}%
      \def\colorgray#1{\color[gray]{#1}}%
      \expandafter\def\csname LTw\endcsname{\color{white}}%
      \expandafter\def\csname LTb\endcsname{\color{black}}%
      \expandafter\def\csname LTa\endcsname{\color{black}}%
      \expandafter\def\csname LT0\endcsname{\color[rgb]{1,0,0}}%
      \expandafter\def\csname LT1\endcsname{\color[rgb]{0,1,0}}%
      \expandafter\def\csname LT2\endcsname{\color[rgb]{0,0,1}}%
      \expandafter\def\csname LT3\endcsname{\color[rgb]{1,0,1}}%
      \expandafter\def\csname LT4\endcsname{\color[rgb]{0,1,1}}%
      \expandafter\def\csname LT5\endcsname{\color[rgb]{1,1,0}}%
      \expandafter\def\csname LT6\endcsname{\color[rgb]{0,0,0}}%
      \expandafter\def\csname LT7\endcsname{\color[rgb]{1,0.3,0}}%
      \expandafter\def\csname LT8\endcsname{\color[rgb]{0.5,0.5,0.5}}%
    \else
      \def\colorrgb#1{\color{black}}%
      \def\colorgray#1{\color[gray]{#1}}%
      \expandafter\def\csname LTw\endcsname{\color{white}}%
      \expandafter\def\csname LTb\endcsname{\color{black}}%
      \expandafter\def\csname LTa\endcsname{\color{black}}%
      \expandafter\def\csname LT0\endcsname{\color{black}}%
      \expandafter\def\csname LT1\endcsname{\color{black}}%
      \expandafter\def\csname LT2\endcsname{\color{black}}%
      \expandafter\def\csname LT3\endcsname{\color{black}}%
      \expandafter\def\csname LT4\endcsname{\color{black}}%
      \expandafter\def\csname LT5\endcsname{\color{black}}%
      \expandafter\def\csname LT6\endcsname{\color{black}}%
      \expandafter\def\csname LT7\endcsname{\color{black}}%
      \expandafter\def\csname LT8\endcsname{\color{black}}%
    \fi
  \fi
    \setlength{\unitlength}{0.0500bp}%
    \ifx\gptboxheight\undefined%
      \newlength{\gptboxheight}%
      \newlength{\gptboxwidth}%
      \newsavebox{\gptboxtext}%
    \fi%
    \setlength{\fboxrule}{0.5pt}%
    \setlength{\fboxsep}{1pt}%
    \definecolor{tbcol}{rgb}{1,1,1}%
\begin{picture}(7646.00,2202.00)%
    \gplgaddtomacro\gplbacktext{%
      \csname LTb\endcsname
      \put(1397,576){\makebox(0,0)[r]{\strut{}$10^{-4}$}}%
      \csname LTb\endcsname
      \put(1397,1138){\makebox(0,0)[r]{\strut{}$10^{-3}$}}%
      \csname LTb\endcsname
      \put(1397,1701){\makebox(0,0)[r]{\strut{}$10^{-2}$}}%
      \csname LTb\endcsname
      \put(1529,356){\makebox(0,0){\strut{}$10^{0}$}}%
      \csname LTb\endcsname
      \put(2051,356){\makebox(0,0){\strut{}$10^{1}$}}%
      \csname LTb\endcsname
      \put(2573,356){\makebox(0,0){\strut{}$10^{2}$}}%
      \csname LTb\endcsname
      \put(3095,356){\makebox(0,0){\strut{}$10^{3}$}}%
    }%
    \gplgaddtomacro\gplfronttext{%
      \csname LTb\endcsname
      \put(836,1223){\makebox(0,0){\strut{}$\widetilde{\lambda}_{n}^{\mathrm{e}}$}}%
      \put(2320,26){\makebox(0,0){\strut{}$n$}}%
      \csname LTb\endcsname
      \put(2320,2200){\makebox(0,0){\strut{}$(a)$}}%
    }%
    \gplgaddtomacro\gplbacktext{%
      \csname LTb\endcsname
      \put(4510,576){\makebox(0,0)[r]{\strut{}$10^{-4}$}}%
      \csname LTb\endcsname
      \put(4510,1138){\makebox(0,0)[r]{\strut{}$10^{-3}$}}%
      \csname LTb\endcsname
      \put(4510,1701){\makebox(0,0)[r]{\strut{}$10^{-2}$}}%
      \csname LTb\endcsname
      \put(4642,356){\makebox(0,0){\strut{}$10^{0}$}}%
      \csname LTb\endcsname
      \put(5164,356){\makebox(0,0){\strut{}$10^{1}$}}%
      \csname LTb\endcsname
      \put(5686,356){\makebox(0,0){\strut{}$10^{2}$}}%
      \csname LTb\endcsname
      \put(6208,356){\makebox(0,0){\strut{}$10^{3}$}}%
      \put(6990,1871){\makebox(0,0)[l]{\strut{}$N'$}}%
    }%
    \gplgaddtomacro\gplfronttext{%
      \csname LTb\endcsname
      \put(3949,1223){\makebox(0,0){\strut{}$\widetilde{\lambda}_{n}^{\mathrm{d},\perp N'}$}}%
      \put(5433,26){\makebox(0,0){\strut{}$n + N'$}}%
      \csname LTb\endcsname
      \put(6899,1651){\makebox(0,0)[r]{\strut{}$0$}}%
      \csname LTb\endcsname
      \put(6899,1431){\makebox(0,0)[r]{\strut{}$10$}}%
      \csname LTb\endcsname
      \put(6899,1211){\makebox(0,0)[r]{\strut{}$50$}}%
      \csname LTb\endcsname
      \put(6899,991){\makebox(0,0)[r]{\strut{}$100$}}%
      \csname LTb\endcsname
      \put(6899,771){\makebox(0,0)[r]{\strut{}$200$}}%
      \csname LTb\endcsname
      \put(5433,2200){\makebox(0,0){\strut{}$(b)$}}%
    }%
    \gplbacktext
    \put(0,0){\includegraphics[width={382.30bp},height={110.10bp}]{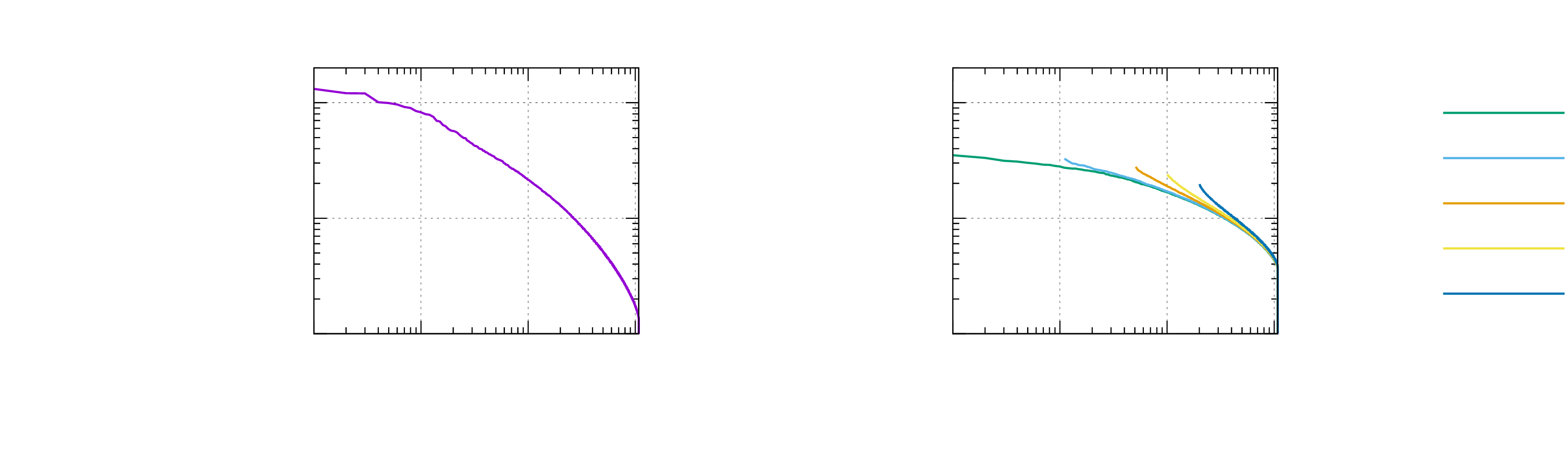}}%
    \gplfronttext
  \end{picture}%
\endgroup

%% file: Figures/convergence_plots.tex
\begingroup
  \makeatletter
  \providecommand\color[2][]{%
    \GenericError{(gnuplot) \space\space\space\@spaces}{%
      Package color not loaded in conjunction with
      terminal option `colourtext'%
    }{See the gnuplot documentation for explanation.%
    }{Either use 'blacktext' in gnuplot or load the package
      color.sty in LaTeX.}%
    \renewcommand\color[2][]{}%
  }%
  \providecommand\includegraphics[2][]{%
    \GenericError{(gnuplot) \space\space\space\@spaces}{%
      Package graphicx or graphics not loaded%
    }{See the gnuplot documentation for explanation.%
    }{The gnuplot epslatex terminal needs graphicx.sty or graphics.sty.}%
    \renewcommand\includegraphics[2][]{}%
  }%
  \providecommand\rotatebox[2]{#2}%
  \@ifundefined{ifGPcolor}{%
    \newif\ifGPcolor
    \GPcolorfalse
  }{}%
  \@ifundefined{ifGPblacktext}{%
    \newif\ifGPblacktext
    \GPblacktexttrue
  }{}%
  \let\gplgaddtomacro\g@addto@macro
  \gdef\gplbacktext{}%
  \gdef\gplfronttext{}%
  \makeatother
  \ifGPblacktext
    \def\colorrgb#1{}%
    \def\colorgray#1{}%
  \else
    \ifGPcolor
      \def\colorrgb#1{\color[rgb]{#1}}%
      \def\colorgray#1{\color[gray]{#1}}%
      \expandafter\def\csname LTw\endcsname{\color{white}}%
      \expandafter\def\csname LTb\endcsname{\color{black}}%
      \expandafter\def\csname LTa\endcsname{\color{black}}%
      \expandafter\def\csname LT0\endcsname{\color[rgb]{1,0,0}}%
      \expandafter\def\csname LT1\endcsname{\color[rgb]{0,1,0}}%
      \expandafter\def\csname LT2\endcsname{\color[rgb]{0,0,1}}%
      \expandafter\def\csname LT3\endcsname{\color[rgb]{1,0,1}}%
      \expandafter\def\csname LT4\endcsname{\color[rgb]{0,1,1}}%
      \expandafter\def\csname LT5\endcsname{\color[rgb]{1,1,0}}%
      \expandafter\def\csname LT6\endcsname{\color[rgb]{0,0,0}}%
      \expandafter\def\csname LT7\endcsname{\color[rgb]{1,0.3,0}}%
      \expandafter\def\csname LT8\endcsname{\color[rgb]{0.5,0.5,0.5}}%
    \else
      \def\colorrgb#1{\color{black}}%
      \def\colorgray#1{\color[gray]{#1}}%
      \expandafter\def\csname LTw\endcsname{\color{white}}%
      \expandafter\def\csname LTb\endcsname{\color{black}}%
      \expandafter\def\csname LTa\endcsname{\color{black}}%
      \expandafter\def\csname LT0\endcsname{\color{black}}%
      \expandafter\def\csname LT1\endcsname{\color{black}}%
      \expandafter\def\csname LT2\endcsname{\color{black}}%
      \expandafter\def\csname LT3\endcsname{\color{black}}%
      \expandafter\def\csname LT4\endcsname{\color{black}}%
      \expandafter\def\csname LT5\endcsname{\color{black}}%
      \expandafter\def\csname LT6\endcsname{\color{black}}%
      \expandafter\def\csname LT7\endcsname{\color{black}}%
      \expandafter\def\csname LT8\endcsname{\color{black}}%
    \fi
  \fi
    \setlength{\unitlength}{0.0500bp}%
    \ifx\gptboxheight\undefined%
      \newlength{\gptboxheight}%
      \newlength{\gptboxwidth}%
      \newsavebox{\gptboxtext}%
    \fi%
    \setlength{\fboxrule}{0.5pt}%
    \setlength{\fboxsep}{1pt}%
    \definecolor{tbcol}{rgb}{1,1,1}%
\begin{picture}(7646.00,4104.00)%
    \gplgaddtomacro\gplbacktext{%
      \csname LTb\endcsname
      \put(656,2500){\makebox(0,0)[r]{\strut{}$0$}}%
      \csname LTb\endcsname
      \put(656,2759){\makebox(0,0)[r]{\strut{}$0.2$}}%
      \csname LTb\endcsname
      \put(656,3018){\makebox(0,0)[r]{\strut{}$0.4$}}%
      \csname LTb\endcsname
      \put(656,3277){\makebox(0,0)[r]{\strut{}$0.6$}}%
      \csname LTb\endcsname
      \put(656,3536){\makebox(0,0)[r]{\strut{}$0.8$}}%
      \csname LTb\endcsname
      \put(656,3795){\makebox(0,0)[r]{\strut{}$1$}}%
      \csname LTb\endcsname
      \put(764,2320){\makebox(0,0){\strut{}}}%
      \csname LTb\endcsname
      \put(1131,2320){\makebox(0,0){\strut{}}}%
      \csname LTb\endcsname
      \put(1499,2320){\makebox(0,0){\strut{}}}%
      \csname LTb\endcsname
      \put(1866,2320){\makebox(0,0){\strut{}}}%
      \csname LTb\endcsname
      \put(2233,2320){\makebox(0,0){\strut{}}}%
    }%
    \gplgaddtomacro\gplfronttext{%
      \csname LTb\endcsname
      \put(161,3147){\rotatebox{-270}{\makebox(0,0){\strut{}$\left\langle T\right\rangle_{\hat{N}}^{\perp N'}$}}}%
      \csname LTb\endcsname
      \put(1474,41){\makebox(0,0)[r]{\strut{}0}}%
      \csname LTb\endcsname
      \put(2401,41){\makebox(0,0)[r]{\strut{}10}}%
      \csname LTb\endcsname
      \put(3328,41){\makebox(0,0)[r]{\strut{}50}}%
      \csname LTb\endcsname
      \put(4255,41){\makebox(0,0)[r]{\strut{}100}}%
      \csname LTb\endcsname
      \put(5182,41){\makebox(0,0)[r]{\strut{}200}}%
      \csname LTb\endcsname
      \put(6109,41){\makebox(0,0)[r]{\strut{}$N$}}%
      \csname LTb\endcsname
      \put(1556,4065){\makebox(0,0){\strut{}($a$)}}%
    }%
    \gplgaddtomacro\gplbacktext{%
      \csname LTb\endcsname
      \put(3286,2500){\makebox(0,0)[r]{\strut{}$0$}}%
      \csname LTb\endcsname
      \put(3286,2759){\makebox(0,0)[r]{\strut{}$0.2$}}%
      \csname LTb\endcsname
      \put(3286,3018){\makebox(0,0)[r]{\strut{}$0.4$}}%
      \csname LTb\endcsname
      \put(3286,3277){\makebox(0,0)[r]{\strut{}$0.6$}}%
      \csname LTb\endcsname
      \put(3286,3536){\makebox(0,0)[r]{\strut{}$0.8$}}%
      \csname LTb\endcsname
      \put(3286,3795){\makebox(0,0)[r]{\strut{}$1$}}%
      \csname LTb\endcsname
      \put(3394,2320){\makebox(0,0){\strut{}}}%
      \csname LTb\endcsname
      \put(3761,2320){\makebox(0,0){\strut{}}}%
      \csname LTb\endcsname
      \put(4128,2320){\makebox(0,0){\strut{}}}%
      \csname LTb\endcsname
      \put(4495,2320){\makebox(0,0){\strut{}}}%
      \csname LTb\endcsname
      \put(4862,2320){\makebox(0,0){\strut{}}}%
    }%
    \gplgaddtomacro\gplfronttext{%
      \csname LTb\endcsname
      \put(2791,3147){\rotatebox{-270}{\makebox(0,0){\strut{}$\left\langle \mathcal{P}\right\rangle_{\hat{N}}^{\perp N'}$}}}%
      \csname LTb\endcsname
      \put(4185,4065){\makebox(0,0){\strut{}($b$)}}%
    }%
    \gplgaddtomacro\gplbacktext{%
      \csname LTb\endcsname
      \put(5915,2500){\makebox(0,0)[r]{\strut{}$0$}}%
      \csname LTb\endcsname
      \put(5915,2759){\makebox(0,0)[r]{\strut{}$0.2$}}%
      \csname LTb\endcsname
      \put(5915,3018){\makebox(0,0)[r]{\strut{}$0.4$}}%
      \csname LTb\endcsname
      \put(5915,3277){\makebox(0,0)[r]{\strut{}$0.6$}}%
      \csname LTb\endcsname
      \put(5915,3536){\makebox(0,0)[r]{\strut{}$0.8$}}%
      \csname LTb\endcsname
      \put(5915,3795){\makebox(0,0)[r]{\strut{}$1$}}%
      \csname LTb\endcsname
      \put(6023,2320){\makebox(0,0){\strut{}}}%
      \csname LTb\endcsname
      \put(6390,2320){\makebox(0,0){\strut{}}}%
      \csname LTb\endcsname
      \put(6757,2320){\makebox(0,0){\strut{}}}%
      \csname LTb\endcsname
      \put(7124,2320){\makebox(0,0){\strut{}}}%
      \csname LTb\endcsname
      \put(7491,2320){\makebox(0,0){\strut{}}}%
    }%
    \gplgaddtomacro\gplfronttext{%
      \csname LTb\endcsname
      \put(5420,3147){\rotatebox{-270}{\makebox(0,0){\strut{}$\left\langle\varepsilon\right\rangle_{\hat{N}}^{\perp N'}$}}}%
      \csname LTb\endcsname
      \put(6814,4065){\makebox(0,0){\strut{}($c$)}}%
    }%
    \gplgaddtomacro\gplbacktext{%
      \csname LTb\endcsname
      \put(656,759){\makebox(0,0)[r]{\strut{}$0$}}%
      \csname LTb\endcsname
      \put(656,1018){\makebox(0,0)[r]{\strut{}$0.05$}}%
      \csname LTb\endcsname
      \put(656,1277){\makebox(0,0)[r]{\strut{}$0.1$}}%
      \csname LTb\endcsname
      \put(656,1536){\makebox(0,0)[r]{\strut{}$0.15$}}%
      \csname LTb\endcsname
      \put(656,1795){\makebox(0,0)[r]{\strut{}$0.2$}}%
      \csname LTb\endcsname
      \put(656,2054){\makebox(0,0)[r]{\strut{}$0.25$}}%
      \csname LTb\endcsname
      \put(764,579){\makebox(0,0){\strut{}$0$}}%
      \csname LTb\endcsname
      \put(1131,579){\makebox(0,0){\strut{}$250$}}%
      \csname LTb\endcsname
      \put(1499,579){\makebox(0,0){\strut{}$500$}}%
      \csname LTb\endcsname
      \put(1866,579){\makebox(0,0){\strut{}$750$}}%
      \csname LTb\endcsname
      \put(2233,579){\makebox(0,0){\strut{}$1000$}}%
      \put(956,41){\makebox(0,0)[l]{\strut{}$N'$:}}%
    }%
    \gplgaddtomacro\gplfronttext{%
      \csname LTb\endcsname
      \put(161,1406){\rotatebox{-270}{\makebox(0,0){\strut{}$\Delta \left\langle T\right\rangle_{\hat{N}}^{\perp N'}$}}}%
      \put(1556,309){\makebox(0,0){\strut{}$\hat{N}$}}%
      \csname LTb\endcsname
      \put(1556,2324){\makebox(0,0){\strut{}($d$)}}%
    }%
    \gplgaddtomacro\gplbacktext{%
      \csname LTb\endcsname
      \put(3286,759){\makebox(0,0)[r]{\strut{}$0$}}%
      \csname LTb\endcsname
      \put(3286,1018){\makebox(0,0)[r]{\strut{}$0.02$}}%
      \csname LTb\endcsname
      \put(3286,1277){\makebox(0,0)[r]{\strut{}$0.04$}}%
      \csname LTb\endcsname
      \put(3286,1536){\makebox(0,0)[r]{\strut{}$0.06$}}%
      \csname LTb\endcsname
      \put(3286,1795){\makebox(0,0)[r]{\strut{}$0.08$}}%
      \csname LTb\endcsname
      \put(3286,2054){\makebox(0,0)[r]{\strut{}$0.1$}}%
      \csname LTb\endcsname
      \put(3394,579){\makebox(0,0){\strut{}$0$}}%
      \csname LTb\endcsname
      \put(3761,579){\makebox(0,0){\strut{}$250$}}%
      \csname LTb\endcsname
      \put(4128,579){\makebox(0,0){\strut{}$500$}}%
      \csname LTb\endcsname
      \put(4495,579){\makebox(0,0){\strut{}$750$}}%
      \csname LTb\endcsname
      \put(4862,579){\makebox(0,0){\strut{}$1000$}}%
    }%
    \gplgaddtomacro\gplfronttext{%
      \csname LTb\endcsname
      \put(2791,1406){\rotatebox{-270}{\makebox(0,0){\strut{}$\Delta \left\langle\mathcal{P}\right\rangle_{\hat{N}}^{\perp N'}$}}}%
      \put(4185,309){\makebox(0,0){\strut{}$\hat{N}$}}%
      \csname LTb\endcsname
      \put(4185,2324){\makebox(0,0){\strut{}($e$)}}%
    }%
    \gplgaddtomacro\gplbacktext{%
      \csname LTb\endcsname
      \put(5915,759){\makebox(0,0)[r]{\strut{}$0$}}%
      \csname LTb\endcsname
      \put(5915,1018){\makebox(0,0)[r]{\strut{}$0.02$}}%
      \csname LTb\endcsname
      \put(5915,1277){\makebox(0,0)[r]{\strut{}$0.04$}}%
      \csname LTb\endcsname
      \put(5915,1536){\makebox(0,0)[r]{\strut{}$0.06$}}%
      \csname LTb\endcsname
      \put(5915,1795){\makebox(0,0)[r]{\strut{}$0.08$}}%
      \csname LTb\endcsname
      \put(5915,2054){\makebox(0,0)[r]{\strut{}$0.1$}}%
      \csname LTb\endcsname
      \put(6023,579){\makebox(0,0){\strut{}$0$}}%
      \csname LTb\endcsname
      \put(6390,579){\makebox(0,0){\strut{}$250$}}%
      \csname LTb\endcsname
      \put(6757,579){\makebox(0,0){\strut{}$500$}}%
      \csname LTb\endcsname
      \put(7124,579){\makebox(0,0){\strut{}$750$}}%
      \csname LTb\endcsname
      \put(7491,579){\makebox(0,0){\strut{}$1000$}}%
    }%
    \gplgaddtomacro\gplfronttext{%
      \csname LTb\endcsname
      \put(5420,1406){\rotatebox{-270}{\makebox(0,0){\strut{}$\Delta \left\langle\varepsilon\right\rangle_{\hat{N}}^{\perp N'}$}}}%
      \put(6814,309){\makebox(0,0){\strut{}$\hat{N}$}}%
      \csname LTb\endcsname
      \put(6814,2324){\makebox(0,0){\strut{}($f$)}}%
    }%
    \gplbacktext
    \put(0,0){\includegraphics[width={382.30bp},height={205.20bp}]{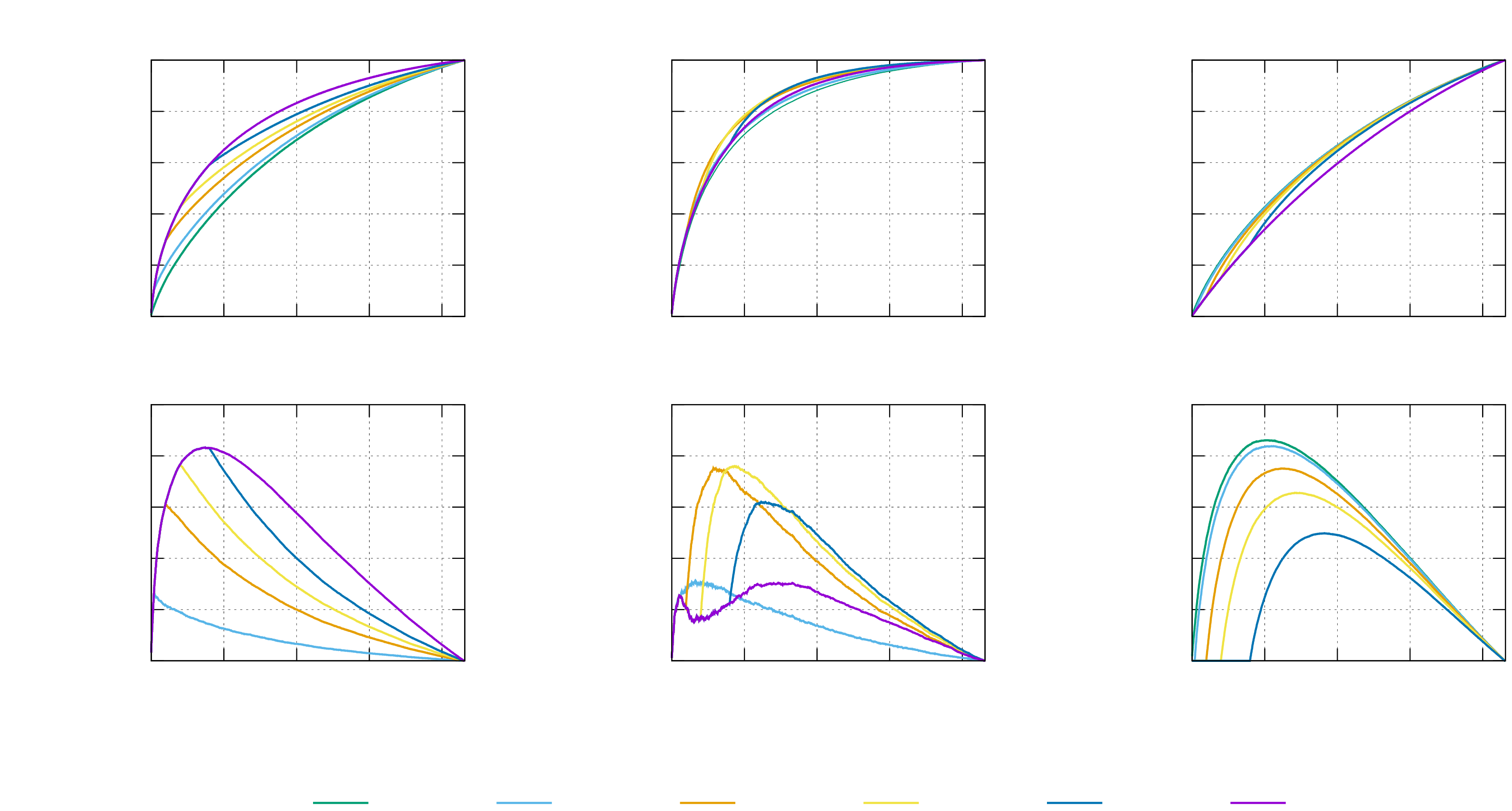}}%
    \gplfronttext
  \end{picture}%
\endgroup

%% file: Figures/all_cascade_DNSinit64amfNEW9-12dt2_c3.tex
\begingroup
  \makeatletter
  \providecommand\color[2][]{%
    \GenericError{(gnuplot) \space\space\space\@spaces}{%
      Package color not loaded in conjunction with
      terminal option `colourtext'%
    }{See the gnuplot documentation for explanation.%
    }{Either use 'blacktext' in gnuplot or load the package
      color.sty in LaTeX.}%
    \renewcommand\color[2][]{}%
  }%
  \providecommand\includegraphics[2][]{%
    \GenericError{(gnuplot) \space\space\space\@spaces}{%
      Package graphicx or graphics not loaded%
    }{See the gnuplot documentation for explanation.%
    }{The gnuplot epslatex terminal needs graphicx.sty or graphics.sty.}%
    \renewcommand\includegraphics[2][]{}%
  }%
  \providecommand\rotatebox[2]{#2}%
  \@ifundefined{ifGPcolor}{%
    \newif\ifGPcolor
    \GPcolortrue
  }{}%
  \@ifundefined{ifGPblacktext}{%
    \newif\ifGPblacktext
    \GPblacktexttrue
  }{}%
  \let\gplgaddtomacro\g@addto@macro
  \gdef\gplbacktext{}%
  \gdef\gplfronttext{}%
  \makeatother
  \ifGPblacktext
    \def\colorrgb#1{}%
    \def\colorgray#1{}%
  \else
    \ifGPcolor
      \def\colorrgb#1{\color[rgb]{#1}}%
      \def\colorgray#1{\color[gray]{#1}}%
      \expandafter\def\csname LTw\endcsname{\color{white}}%
      \expandafter\def\csname LTb\endcsname{\color{black}}%
      \expandafter\def\csname LTa\endcsname{\color{black}}%
      \expandafter\def\csname LT0\endcsname{\color[rgb]{1,0,0}}%
      \expandafter\def\csname LT1\endcsname{\color[rgb]{0,1,0}}%
      \expandafter\def\csname LT2\endcsname{\color[rgb]{0,0,1}}%
      \expandafter\def\csname LT3\endcsname{\color[rgb]{1,0,1}}%
      \expandafter\def\csname LT4\endcsname{\color[rgb]{0,1,1}}%
      \expandafter\def\csname LT5\endcsname{\color[rgb]{1,1,0}}%
      \expandafter\def\csname LT6\endcsname{\color[rgb]{0,0,0}}%
      \expandafter\def\csname LT7\endcsname{\color[rgb]{1,0.3,0}}%
      \expandafter\def\csname LT8\endcsname{\color[rgb]{0.5,0.5,0.5}}%
    \else
      \def\colorrgb#1{\color{black}}%
      \def\colorgray#1{\color[gray]{#1}}%
      \expandafter\def\csname LTw\endcsname{\color{white}}%
      \expandafter\def\csname LTb\endcsname{\color{black}}%
      \expandafter\def\csname LTa\endcsname{\color{black}}%
      \expandafter\def\csname LT0\endcsname{\color{black}}%
      \expandafter\def\csname LT1\endcsname{\color{black}}%
      \expandafter\def\csname LT2\endcsname{\color{black}}%
      \expandafter\def\csname LT3\endcsname{\color{black}}%
      \expandafter\def\csname LT4\endcsname{\color{black}}%
      \expandafter\def\csname LT5\endcsname{\color{black}}%
      \expandafter\def\csname LT6\endcsname{\color{black}}%
      \expandafter\def\csname LT7\endcsname{\color{black}}%
      \expandafter\def\csname LT8\endcsname{\color{black}}%
    \fi
  \fi
    \setlength{\unitlength}{0.0500bp}%
    \ifx\gptboxheight\undefined%
      \newlength{\gptboxheight}%
      \newlength{\gptboxwidth}%
      \newsavebox{\gptboxtext}%
    \fi%
    \setlength{\fboxrule}{0.5pt}%
    \setlength{\fboxsep}{1pt}%
    \definecolor{tbcol}{rgb}{1,1,1}%
\begin{picture}(7646.00,5542.00)%
    \gplgaddtomacro\gplbacktext{%
      \csname LTb\endcsname
      \put(784,4225){\makebox(0,0)[r]{\strut{}$10^{-1}$}}%
      \csname LTb\endcsname
      \put(784,4514){\makebox(0,0)[r]{\strut{}$10^{0}$}}%
      \csname LTb\endcsname
      \put(784,4803){\makebox(0,0)[r]{\strut{}$10^{1}$}}%
      \csname LTb\endcsname
      \put(784,5092){\makebox(0,0)[r]{\strut{}$10^{2}$}}%
      \csname LTb\endcsname
      \put(955,3726){\makebox(0,0){\strut{}}}%
      \csname LTb\endcsname
      \put(1272,3726){\makebox(0,0){\strut{}}}%
      \csname LTb\endcsname
      \put(1589,3726){\makebox(0,0){\strut{}}}%
      \csname LTb\endcsname
      \put(1905,3726){\makebox(0,0){\strut{}}}%
      \csname LTb\endcsname
      \put(2222,3726){\makebox(0,0){\strut{}}}%
      \csname LTb\endcsname
      \put(2539,3726){\makebox(0,0){\strut{}}}%
    }%
    \gplgaddtomacro\gplfronttext{%
      \csname LTb\endcsname
      \put(217,4616){\makebox(0,0){\strut{}$y^+$}}%
      \csname LTb\endcsname
      \put(1747,5443){\makebox(0,0){\strut{}$(a)$}}%
    }%
    \gplgaddtomacro\gplbacktext{%
      \csname LTb\endcsname
      \put(2802,4225){\makebox(0,0)[r]{\strut{}}}%
      \csname LTb\endcsname
      \put(2802,4514){\makebox(0,0)[r]{\strut{}}}%
      \csname LTb\endcsname
      \put(2802,4803){\makebox(0,0)[r]{\strut{}}}%
      \csname LTb\endcsname
      \put(2802,5092){\makebox(0,0)[r]{\strut{}}}%
      \csname LTb\endcsname
      \put(2973,3726){\makebox(0,0){\strut{}}}%
      \csname LTb\endcsname
      \put(3290,3726){\makebox(0,0){\strut{}}}%
      \csname LTb\endcsname
      \put(3607,3726){\makebox(0,0){\strut{}}}%
      \csname LTb\endcsname
      \put(3923,3726){\makebox(0,0){\strut{}}}%
      \csname LTb\endcsname
      \put(4240,3726){\makebox(0,0){\strut{}}}%
      \csname LTb\endcsname
      \put(4557,3726){\makebox(0,0){\strut{}}}%
    }%
    \gplgaddtomacro\gplfronttext{%
      \csname LTb\endcsname
      \put(3765,5443){\makebox(0,0){\strut{}$(b)$}}%
    }%
    \gplgaddtomacro\gplbacktext{%
      \csname LTb\endcsname
      \put(4820,4225){\makebox(0,0)[r]{\strut{}}}%
      \csname LTb\endcsname
      \put(4820,4514){\makebox(0,0)[r]{\strut{}}}%
      \csname LTb\endcsname
      \put(4820,4803){\makebox(0,0)[r]{\strut{}}}%
      \csname LTb\endcsname
      \put(4820,5092){\makebox(0,0)[r]{\strut{}}}%
      \csname LTb\endcsname
      \put(4991,3726){\makebox(0,0){\strut{}}}%
      \csname LTb\endcsname
      \put(5308,3726){\makebox(0,0){\strut{}}}%
      \csname LTb\endcsname
      \put(5624,3726){\makebox(0,0){\strut{}}}%
      \csname LTb\endcsname
      \put(5941,3726){\makebox(0,0){\strut{}}}%
      \csname LTb\endcsname
      \put(6257,3726){\makebox(0,0){\strut{}}}%
      \csname LTb\endcsname
      \put(6574,3726){\makebox(0,0){\strut{}}}%
    }%
    \gplgaddtomacro\gplfronttext{%
      \csname LTb\endcsname
      \put(5782,5443){\makebox(0,0){\strut{}$(c)$}}%
    }%
    \gplgaddtomacro\gplbacktext{%
      \csname LTb\endcsname
      \put(784,2573){\makebox(0,0)[r]{\strut{}$10^{-1}$}}%
      \csname LTb\endcsname
      \put(784,2862){\makebox(0,0)[r]{\strut{}$10^{0}$}}%
      \csname LTb\endcsname
      \put(784,3152){\makebox(0,0)[r]{\strut{}$10^{1}$}}%
      \csname LTb\endcsname
      \put(784,3441){\makebox(0,0)[r]{\strut{}$10^{2}$}}%
      \csname LTb\endcsname
      \put(955,2074){\makebox(0,0){\strut{}}}%
      \csname LTb\endcsname
      \put(1272,2074){\makebox(0,0){\strut{}}}%
      \csname LTb\endcsname
      \put(1589,2074){\makebox(0,0){\strut{}}}%
      \csname LTb\endcsname
      \put(1905,2074){\makebox(0,0){\strut{}}}%
      \csname LTb\endcsname
      \put(2222,2074){\makebox(0,0){\strut{}}}%
      \csname LTb\endcsname
      \put(2539,2074){\makebox(0,0){\strut{}}}%
    }%
    \gplgaddtomacro\gplfronttext{%
      \csname LTb\endcsname
      \put(217,2964){\makebox(0,0){\strut{}$y^+$}}%
      \csname LTb\endcsname
      \put(1747,3792){\makebox(0,0){\strut{}$(d)$}}%
    }%
    \gplgaddtomacro\gplbacktext{%
      \csname LTb\endcsname
      \put(2802,2573){\makebox(0,0)[r]{\strut{}}}%
      \csname LTb\endcsname
      \put(2802,2862){\makebox(0,0)[r]{\strut{}}}%
      \csname LTb\endcsname
      \put(2802,3152){\makebox(0,0)[r]{\strut{}}}%
      \csname LTb\endcsname
      \put(2802,3441){\makebox(0,0)[r]{\strut{}}}%
      \csname LTb\endcsname
      \put(2973,2074){\makebox(0,0){\strut{}}}%
      \csname LTb\endcsname
      \put(3290,2074){\makebox(0,0){\strut{}}}%
      \csname LTb\endcsname
      \put(3607,2074){\makebox(0,0){\strut{}}}%
      \csname LTb\endcsname
      \put(3923,2074){\makebox(0,0){\strut{}}}%
      \csname LTb\endcsname
      \put(4240,2074){\makebox(0,0){\strut{}}}%
      \csname LTb\endcsname
      \put(4557,2074){\makebox(0,0){\strut{}}}%
    }%
    \gplgaddtomacro\gplfronttext{%
      \csname LTb\endcsname
      \put(3765,3792){\makebox(0,0){\strut{}$(e)$}}%
    }%
    \gplgaddtomacro\gplbacktext{%
      \csname LTb\endcsname
      \put(4820,2573){\makebox(0,0)[r]{\strut{}}}%
      \csname LTb\endcsname
      \put(4820,2862){\makebox(0,0)[r]{\strut{}}}%
      \csname LTb\endcsname
      \put(4820,3152){\makebox(0,0)[r]{\strut{}}}%
      \csname LTb\endcsname
      \put(4820,3441){\makebox(0,0)[r]{\strut{}}}%
      \csname LTb\endcsname
      \put(4991,2074){\makebox(0,0){\strut{}}}%
      \csname LTb\endcsname
      \put(5308,2074){\makebox(0,0){\strut{}}}%
      \csname LTb\endcsname
      \put(5624,2074){\makebox(0,0){\strut{}}}%
      \csname LTb\endcsname
      \put(5941,2074){\makebox(0,0){\strut{}}}%
      \csname LTb\endcsname
      \put(6257,2074){\makebox(0,0){\strut{}}}%
      \csname LTb\endcsname
      \put(6574,2074){\makebox(0,0){\strut{}}}%
    }%
    \gplgaddtomacro\gplfronttext{%
      \csname LTb\endcsname
      \put(5782,3792){\makebox(0,0){\strut{}$(f)$}}%
    }%
    \gplgaddtomacro\gplbacktext{%
      \csname LTb\endcsname
      \put(784,921){\makebox(0,0)[r]{\strut{}$10^{-1}$}}%
      \csname LTb\endcsname
      \put(784,1210){\makebox(0,0)[r]{\strut{}$10^{0}$}}%
      \csname LTb\endcsname
      \put(784,1500){\makebox(0,0)[r]{\strut{}$10^{1}$}}%
      \csname LTb\endcsname
      \put(784,1789){\makebox(0,0)[r]{\strut{}$10^{2}$}}%
      \csname LTb\endcsname
      \put(955,422){\makebox(0,0){\strut{}$0$}}%
      \csname LTb\endcsname
      \put(1272,422){\makebox(0,0){\strut{}$1$}}%
      \csname LTb\endcsname
      \put(1589,422){\makebox(0,0){\strut{}$2$}}%
      \csname LTb\endcsname
      \put(1905,422){\makebox(0,0){\strut{}$3$}}%
      \csname LTb\endcsname
      \put(2222,422){\makebox(0,0){\strut{}$4$}}%
      \csname LTb\endcsname
      \put(2539,422){\makebox(0,0){\strut{}$5$}}%
    }%
    \gplgaddtomacro\gplfronttext{%
      \csname LTb\endcsname
      \put(217,1312){\makebox(0,0){\strut{}$y^+$}}%
      \put(1747,152){\makebox(0,0){\strut{}$\left(\left\langle T\right\rangle_{\hat{N}}^{\perp N'}\right)^+$}}%
      \csname LTb\endcsname
      \put(1747,2140){\makebox(0,0){\strut{}$(g)$}}%
    }%
    \gplgaddtomacro\gplbacktext{%
      \csname LTb\endcsname
      \put(2802,921){\makebox(0,0)[r]{\strut{}}}%
      \csname LTb\endcsname
      \put(2802,1210){\makebox(0,0)[r]{\strut{}}}%
      \csname LTb\endcsname
      \put(2802,1500){\makebox(0,0)[r]{\strut{}}}%
      \csname LTb\endcsname
      \put(2802,1789){\makebox(0,0)[r]{\strut{}}}%
      \csname LTb\endcsname
      \put(2973,422){\makebox(0,0){\strut{}$0$}}%
      \csname LTb\endcsname
      \put(3290,422){\makebox(0,0){\strut{}$0.05$}}%
      \csname LTb\endcsname
      \put(3607,422){\makebox(0,0){\strut{}$0.1$}}%
      \csname LTb\endcsname
      \put(3923,422){\makebox(0,0){\strut{}$0.15$}}%
      \csname LTb\endcsname
      \put(4240,422){\makebox(0,0){\strut{}$0.2$}}%
      \csname LTb\endcsname
      \put(4557,422){\makebox(0,0){\strut{}$0.25$}}%
    }%
    \gplgaddtomacro\gplfronttext{%
      \csname LTb\endcsname
      \put(3765,152){\makebox(0,0){\strut{}$\left(\left\langle \mathcal{P}\right\rangle_{\hat{N}}^{\perp N'}\right)^+$}}%
      \csname LTb\endcsname
      \put(3765,2140){\makebox(0,0){\strut{}$(h)$}}%
    }%
    \gplgaddtomacro\gplbacktext{%
      \csname LTb\endcsname
      \put(4820,921){\makebox(0,0)[r]{\strut{}}}%
      \csname LTb\endcsname
      \put(4820,1210){\makebox(0,0)[r]{\strut{}}}%
      \csname LTb\endcsname
      \put(4820,1500){\makebox(0,0)[r]{\strut{}}}%
      \csname LTb\endcsname
      \put(4820,1789){\makebox(0,0)[r]{\strut{}}}%
      \csname LTb\endcsname
      \put(4991,422){\makebox(0,0){\strut{}$0$}}%
      \csname LTb\endcsname
      \put(5308,422){\makebox(0,0){\strut{}$0.05$}}%
      \csname LTb\endcsname
      \put(5624,422){\makebox(0,0){\strut{}$0.1$}}%
      \csname LTb\endcsname
      \put(5941,422){\makebox(0,0){\strut{}$0.15$}}%
      \csname LTb\endcsname
      \put(6257,422){\makebox(0,0){\strut{}$0.2$}}%
      \csname LTb\endcsname
      \put(6574,422){\makebox(0,0){\strut{}$0.25$}}%
    }%
    \gplgaddtomacro\gplfronttext{%
      \csname LTb\endcsname
      \put(5782,152){\makebox(0,0){\strut{}$\left(\left\langle\varepsilon\right\rangle_{\hat{N}}^{\perp N'}\right)^+$}}%
      \csname LTb\endcsname
      \put(7232,665){\makebox(0,0)[l]{\strut{}$0$}}%
      \put(7232,1518){\makebox(0,0)[l]{\strut{}$200$}}%
      \put(7232,2371){\makebox(0,0)[l]{\strut{}$400$}}%
      \put(7232,3224){\makebox(0,0)[l]{\strut{}$600$}}%
      \put(7232,4077){\makebox(0,0)[l]{\strut{}$800$}}%
      \put(7232,4931){\makebox(0,0)[l]{\strut{}$1000$}}%
      \put(7394,5394){\makebox(0,0){\strut{}$\hat{N}$}}%
      \put(5782,2140){\makebox(0,0){\strut{}$(i)$}}%
    }%
    \gplbacktext
    \put(0,0){\includegraphics[width={382.30bp},height={277.10bp}]{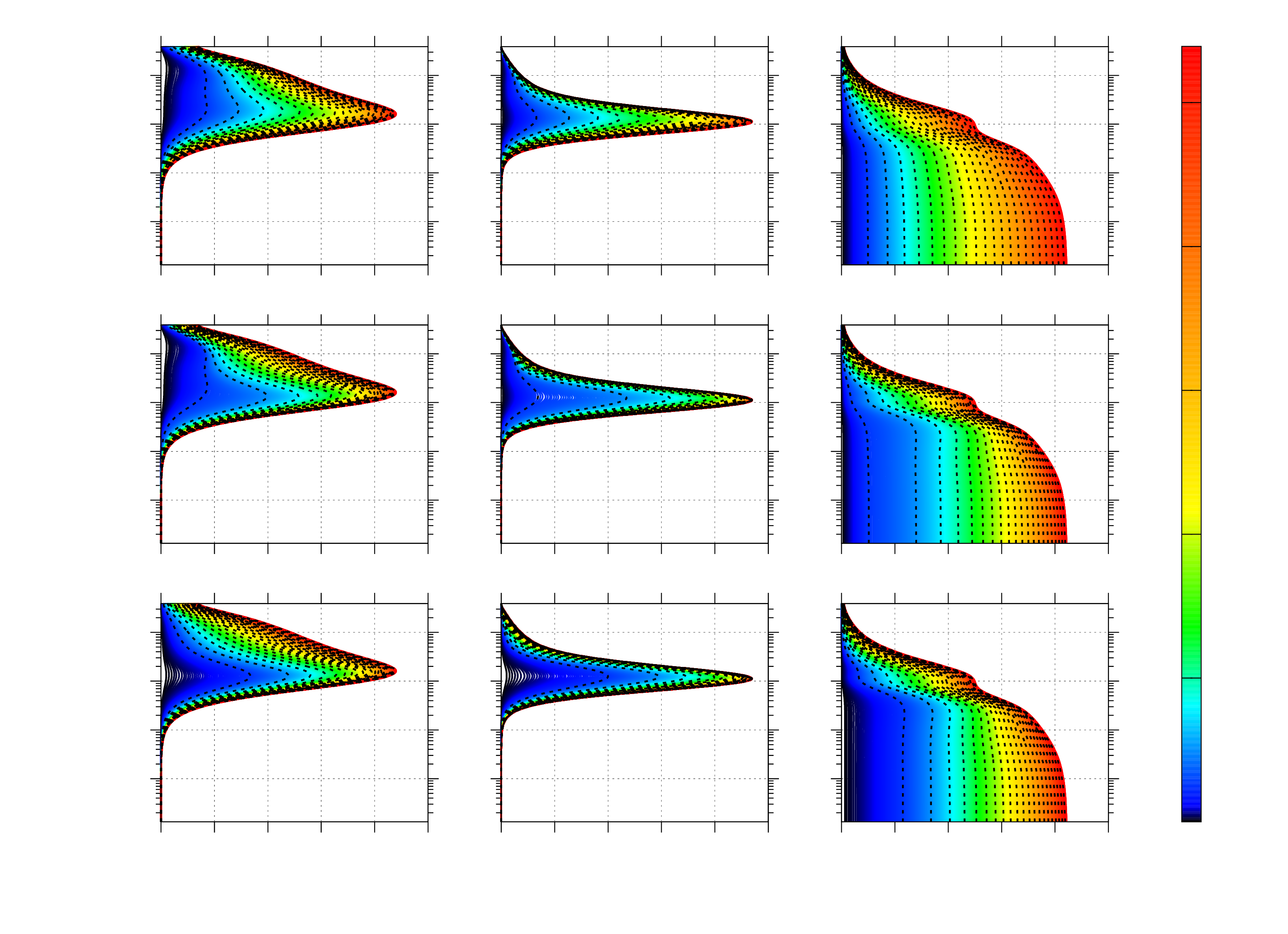}}%
    \gplfronttext
  \end{picture}%
\endgroup